\documentclass[twocolumn, pra, superscriptaddress,notitlepage]{revtex4-1}
\usepackage{graphicx}
\usepackage{physics}
\usepackage{epsfig}
\usepackage{color}
\usepackage{xspace}
\usepackage{array}
\usepackage{amsthm,amsmath,amssymb,mathrsfs}
\usepackage{amsfonts}
\usepackage{ulem}
\usepackage{bm}
\usepackage[dvipsnames]{xcolor}
\usepackage{colortbl}
\usepackage[breaklinks,color links,linkcolor=blue,citecolor=blue,urlcolor=blue]{hyperref}
\makeatletter

\newcommand{\Rmnum}[1]{\expandafter\@slowromancap\romannumeral #1@}
\makeatother

\makeatletter
\let\Hy@backout\@gobble
\makeatother
\begin{document}
\title{Generalized Spectral Form Factor in Random Matrix Theory}

\author{Zhiyang Wei}
\affiliation{MOE Key Laboratory for Nonequilibrium Synthesis and Modulation of Condensed Matter, Shaanxi Province Key Laboratory of Quantum
Information and Quantum Optoelectronic Devices, School of Physics, Xi’an Jiaotong University, Xi’an 710049, China}

\author{Chengming Tan}
\affiliation{Hefei National Research Center for Physical Sciences at the Microscale and School of Physics,
University of Science and Technology of China, Hefei 230026, China}

\author{Ren Zhang}
\email{ renzhang@xjtu.edu.cn}
\affiliation{MOE Key Laboratory for Nonequilibrium Synthesis and Modulation of Condensed Matter, Shaanxi Province Key Laboratory of Quantum
Information and Quantum Optoelectronic Devices, School of Physics, Xi’an Jiaotong University, Xi’an 710049, China}
\affiliation{Hefei National Laboratory, Hefei, 230088, China}
\date{\today}

\begin{abstract}
The spectral form factor (SFF) plays a crucial role in revealing the statistical properties of energy level distributions in complex systems. It is one of the tools to diagnose quantum chaos and unravel the universal dynamics therein. The definition of SFF in most literature only encapsulates the two-level correlation. In this manuscript, we extend the definition of SSF to include the high-order correlation. Specifically, we introduce the standard deviation of energy levels to define correlation functions, from which the generalized spectral form factor (GSFF) can be obtained by Fourier transforms. GSFF provides a more comprehensive knowledge of the dynamics of chaotic systems. Using random matrices as examples, we demonstrate new dynamics features that are encoded in GSFF. Remarkably, the GSFF is complex, and the real and imaginary parts exhibit universal dynamics. For instance, in the two-level correlated case, the real part of GSFF shows a dip-ramp-plateau structure akin to the conventional counterpart, and the imaginary part for different system sizes converges in the long time limit. For the two-level GSFF, the analytical forms of the real part are obtained and consistent with numerical results. The results of the imaginary part are obtained by numerical calculation. Similar analyses are extended to three-level GSFF.
\end{abstract}

\maketitle
\section{introduction}
The spectral form factor (SFF) is a powerful tool for characterizing and analyzing the statistical behavior of energy levels in various physical systems, ranging from atomic nuclei~\cite{BREZIN1993613, BREZIN1996697, Papen2007Coll,gomez2011many} to quantum chaotic systems~\cite{gomez2011many,kot1997Q,brezin1997spectral,links2003algebraic,turek2005Semi,uller2005Pot,sven2006Q,bertini2018exact,chan2018spectral,kos2018Many,roy2022Spectral,chen2018universal,del2018decay,liu2018spectral,chenu2019work,fri2019Spectral,gai2019Spec,cao2022prob,winer2022Losch,bal2022Q,buivi2022quantum,ste1984Q}. 
It measures the fluctuations of the density of states in a complex system and reveals universal features of quantum chaotic systems~\cite{haake1991quantum}, such as level repulsion~\cite{bohi1984charact, CAURIER1989387}, random matrix statistics~\cite{bohigas1991random, Andreev1996QC}, and quantum ergodicity~\cite{ste1984Q, David2015QE, MADHOK2018189}. One of the pioneering developments that establish the spectrum statistical properties is known as the Wigner-Dyson statistics~\cite{wigner1951,wigner1993,dyson1_2004,dyson2_2004,wigner1967random}. It revealed that energy levels in complex systems do not merely follow a simple random pattern but instead exhibit correlations and repulsions, resembling the behavior of eigenvalues of random matrices~\cite{wigner1967random,guhr1998random,mehta2004random}. Therefore, random matrix theory (RMT) is widely used in the study of spectrum statistics and serves as a versatile tool for replicating energy level distributions in complex systems~\cite{wigner1967random,guhr1998random,mehta2004random}.

The SFF finds diverse applications across various disciplines. In condensed matter physics, it has been used to diagnose the quantum phase transitions and the level statistics of many-body systems, such as spin chains, disordered systems, and topological insulators~\cite{rabson2004cross,joshi2022prob,kos2018Many,chan2018,chan2018spectral,bertini2018exact,janez2020,nive2020,liao2020many,sier2020thou,winer2022HT,roy2022Spectral,winer2023Emergent,winer2020ramp,roy2020random,ward2021topo,sarkar2023spectral, Barney_2023}. For instance, the SFF is employed to analyze a hydrodynamic system with a sound pole in a finite volume cavity, revealing an association of the logarithm of the hydrodynamic enhancement with a quantum particle moving in the selfsame cavity~\cite{winer2023Emergent}. In the quadratic Sachdev-Ye-Kitaev model, the SFF features an exponential ramp, which contrasts with the linear slopes in chaotic models~\cite{winer2020ramp}. In Floquet fermionic chains with long-range two-particle interactions, SSF precisely follows the prediction of RMT in long chains and timescales surpassing the Thouless time scales with the system size ~\cite{roy2020random}. In quantum field theory and holography, the SFF serves as a tool to probe quantum chaos therein and information scrambling of strongly coupled systems, such as conformal field theories, gauge theories, and black holes~\cite{buivi2022quantum,caceres2022spectral,belin2022generalized,de2019spectral,cotler2017black,li2017ssym,kudler2020CFT,easther2006random,chen2022spectral,choi2023supersymmetric}. For example, the SFF of supersymmetric Yang-Mills theory in four dimensions exhibits consistency with that of a Euclidean black hole in AdS$_5$~\cite{choi2023supersymmetric}, which provides a geometrical interpretation of the ramp behavior. In addition, there are many studies in the field of quantum information. A major modification to the bipartite entanglement entropy at large sub-system size depends only on the theoretical particle spectrum~\cite{cardy2008form}; The concept of SFF can be naturally extended to a subsystem via pseudo--entropy ~\cite{goto2021subregion}; Also, the study of SSF  have been extended to open systems or non-Hermitian physics~\cite{li2021Spectral, Xu2021TD, Kos2021chaos, Corne2022Spectral,zhou2023universal, Kawa2023Dqh,matsoukas2023nohermitian,matsoukas2023unitarity,matsoukas2023quantum,roccati2023diagnosing}. 

As of now, most of the research focuses on the SFF with a two-level correlation, which encapsulates the energy level spacing, denoting the intervals between two energy levels. To our best knowledge, a few works extend SFF to the case of higher even-order correlation, such as four-point correlation~\cite{liu2018spectral,cardella2021late}, and six-point correlation~\cite{gross_all_2017}. A universal definition for the case of arbitrary-order correlation is still lacking. In this manuscript, we employ the standard deviation of energy levels to redefine the correlation function and obtain the generalized spectral form factor (GSFF) through a Fourier transform. The GSFF elucidates the characteristics of high-order fluctuations in the density of energy states, which manifest themselves in specific structural properties in the time domain. By studying the GSFF in RMT, we find that it is a complex quantity. For the case of two-level correlation, we obtain the same dip-ramp-plateau structure as that in the conventional SFF~\cite{Saad2018ASR,ghar2018onset, Winer2020ER}. The imaginary part of the GSFF also presents a universal behavior akin to its real counterpart.  These findings are based on an analytical derivation and numerically validated by adopting the Gaussian ensembles. We further find the validity of the aforementioned results can be extended to cases with higher-level correlations, which broadens the scope of our analysis.

The content of this manuscript is organized as follows: In Sec. \ref{sec:2GSFF}, we briefly introduce the Gaussian ensemble and redefine the correlation function to obtain GSFF. In Sec. \ref{sec:3GSFF}, we calculate the GSFF associated with two energy levels in GUE and demonstrate the universal behavior of the real and imaginary parts, respectively. In Sec. \ref{sec:4GSFF}, we calculated the GSFF associated with three energy levels correlation, demonstrating the universal behavior of real and imaginary parts. Lastly, we summarize our results in Sec. \ref{summary}.

\section{Definition of generalized spectral form factor}
\label{sec:2GSFF}
\subsection{Gaussian ensemble}
For the sake of the following discussion, let us start with a brief introduction to the Gaussian ensemble which is a pivotal concept in the realm of RMT, owing its origins to Eugene Wigner's groundbreaking research in the late 1950s~\cite{wigner1951,wigner1967random}. By investigating the distribution of random matrices, Wigner obtained the semicircle law governing the spectrum distribution of Gaussian ensembles. Within RMT, diverse Hamiltonians are crafted, with their matrix elements following Gaussian distributions. These ensembles can be categorized based on matrix element characteristics (real, complex, or quaternion) and matrix symmetry, resulting in the Gaussian orthogonal ensemble (GOE), Gaussian unitary ensemble (GUE), and Gaussian symplectic ensemble (GSE), collectively known as GXE. These ensembles show different Dyson's index, denoted as $\beta$, where GOE corresponds to $\beta = 1$, GUE to $\beta = 2$, and GSE to $\beta = 4$~\cite{wigner1967random,guhr1998random,mehta2004random}. 

The applications of these ensembles have extended to the realm of quantum physics, with GOE applying to systems possessing time-reversal symmetry and an even number or no 1/2 spin particles. In contrast, GUE is associated with systems in the absence of time-reversal symmetry, featuring random complex symmetric Hermitian matrices. GSE finds relevance in systems characterized by time-reversal symmetry and comprising an odd number of 1/2 spin particles, involving random real quaternion elements symmetric Hermitian matrices. These concepts, stemming from Wigner's pioneering work, play an important role in the exploration of quantum systems and statistical physics~\cite{wigner1967random,guhr1998random,mehta2004random}. In this manuscript, we will use GUE as an example to demonstrate our results.

\subsection{Spectral correlation function and generalized spectral form factor}
Taking the Gaussian ensemble as an example, we show the generalization of the spectral correlation function and SFF. The Gaussian ensemble comprises a set of random matrices denoted as $\left\lbrace M_{N \times N} \right\rbrace$, each having $N$ eigenvalues labeled as $\lambda_n \left(n = 1, 2, \ldots, N\right)$. The distribution of these eigenvalues is determined by a joint probability density function (JPDF), denoted as $P_N \left(\lambda_1, \ldots, \lambda_N\right)$, which can be derived from the Gaussian distribution. The JPDF represents the probability of finding an energy level at each position along the energy axis defined by $\left\lbrace \lambda_1, \ldots, \lambda_N\right\rbrace$. According to RMT, for some complex physical systems, we can employ random matrices to replicate the energy level distribution of the system~\cite{wigner1967random,guhr1998random,mehta2004random}. The Hamiltonian is written as $H = E \times M_{N \times N}$, where $E$ is the characteristic energy of a particular system, and serves as the energy unit. The eigenenergies of the system are denoted by $\lambda_n \left(n = 1, 2, \ldots, N\right)$. 
It assumes that each energy level is equivalent, implying that the JPDF $P_N \left(\lambda_1, \ldots, \lambda_N\right)$ remains invariant under parameter rearrangements. 

Using the JPDF, we can explore the characteristics of these energy levels. In this manuscript, we employ the standard deviation of the levels to define the correlation among $k$ levels, which is referred to as the $k$-point correlation function and can be expressed as
\begin{equation}
c^{(k)} \left(\epsilon\right) = \int \frac{P[{\bm \lambda}]}{N^k}   \sum_{i=1}^N \delta \left[\epsilon - \sigma_{m} \left(\lambda_{m_{1}}, \ldots, \lambda_{m_{k}}\right)\right] D{\bm \lambda},
\label{1}
\end{equation}
where $k$ is a positive integer smaller than $N$, $P[{\bm \lambda}] = P(\lambda_1, \ldots, \lambda_N)$ represents the JPDF of the $N$ energy levels, $D{\bm \lambda} = d\lambda_1 \ldots d\lambda_N$ denotes the $N$-dimensional integral measures, and $\sigma_{m} \left(\lambda_{m_{1}}, \ldots, \lambda_{m_{k}}\right),m_1,\ldots,m_k \in \left\lbrace 1,2,\ldots, N \right\rbrace $ calculates the standard deviation among any $k$ energy levels. Here, we would like to emphasize that our definition can also be implemented in the case that $k>2$. This is a generalization of the conventional two-point correlation function. It is clear that the large $\epsilon$ contribution of $c^{(k)}(\epsilon)$ comes from the energy configuration that is far-away separated, while the small $\epsilon$ contribution comes from the configurations that are close to each other.

The correlation function defined in Eq.(\ref{1}) can be numerically computed via the ensemble average of a random matrix ensemble. To be specific, we randomly chose $k$ energy levels from the whole spectrum of dimension $N$. So, there are $N^{k}$ configurations for each Hamiltonian in the ensemble. Then we implement an ensemble average.  As a result, the explicit expression of $c^{(k)} \left(\epsilon\right)$ can be requested as
\begin{equation}
c^{(k)} \left(\epsilon\right)= \left\langle \frac{1}{N^k} \sum_{m=1}^{N^{k}} \delta \left(\epsilon - \sigma_{m} \left(\lambda_{m_{1}}, \ldots, \lambda_{m_{k}}\right)\right) \right\rangle_{\text{em}},
\label{2}
\end{equation}
where $\left\langle .\right\rangle_{\text{em}}$ denotes the ensemble average. When $k=1$, $c^{(1)} \left(\epsilon\right) = 1$, which is reasonable. When $k=2$, $\sigma_m \left(\lambda_{m_1}, \lambda_{m_2}\right) = \sigma \left(\lambda_i, \lambda_j\right) = \frac{\left| \lambda_i - \lambda_j \right|}{2}, i, j \in \left\lbrace 1,2,\ldots, N\right\rbrace$, and we obtain the statistical properties of the absolute energy level spacing, which will be further discussed in subsequent sections. Importantly, this framework can be straightforwardly extended to the higher-order correlation.

The generalization of SFF can be obtained by a Fourier transformation of the correlation function defined in Eq.(\ref{1}),
\begin{equation}
\begin{aligned}
 g^{(k)}\left( t\right) = \int c^{(k)}\left( \epsilon \right)  {\rm e}^{-i \epsilon t} d\epsilon .
 \label{3}
\end{aligned}
\end{equation}
For conciseness, we take $\hbar=1$ throughout this manuscript. Employing the GSFF, we gain more insight into the discrete nature of the energy level distribution. The short-time behavior of $ g^{(k)}\left( t\right)$ originates from the energy configuration that is far-away separated. In contrast, the long-time behavior is dominated by the configurations that are close to each other.
 
Compared with previous research of SFF, in the case of two energy levels correlation, GSFF is similar to SFF which is conventionally defined by the difference between energy levels. In contrast, GSFF features an imaginary part showing universal behavior, as we will discuss later. In literature, the high-order energy-level correlation has been considered, but only for even order cases, such as four order~\cite{liu2018spectral}.
Our definition of GSFF is different from the previous ones. The GSFF in our manuscript can be extended to any order correlation, not only even-order correlation but also odd-order correlation.

\section{Two-level Generalized spectral form factor in GUE}
\label{sec:3GSFF}
\subsection{Calculation}
Let us  start with the two-level scenario, where the correlation between energy levels is encapsulated in the two-point correlation function, which is defined as:
\begin{equation}
c^{(2)}\left( \epsilon\right) = \frac{1}{N^2}\int D{\bm\lambda} P[{\bm\lambda}] \sum_{i, j=1}^N \delta\left[\epsilon- \sigma \left(\lambda_i,\lambda_j\right)\right] ,
\label{4}
\end{equation}
where $\sigma(\lambda_i, \lambda_j)$ represents the standard deviation between any pair of energy levels, which can be expressed as the absolute difference between $\lambda_{i}$ and $\lambda_{j}$, i.e., $\sigma(\lambda_i, \lambda_j)=|\lambda_i- \lambda_j|/2$. The definition of integral measure $D{\bm \lambda}$ is the same as that in Eq.(\ref{1}), and the integral interval for each component is $[-2,2]$. Consequently, the two-level GSFF is formulated as follows:
\begin{align}
g^{(2)}\left( t\right) &=\int c^{(2)}\left( \epsilon \right) {\rm e}^{-i \epsilon t} d\epsilon\nonumber\\
& =\frac{1}{N^2} \int d \epsilon\int D{\bm\lambda} P[{\bm\lambda}] \sum_{i, j=1}^N \delta \left[\epsilon-|\lambda_i- \lambda_j|/2\right]  {\rm e}^{-i \epsilon t}\nonumber  \\
& =\frac{1}{N^2} \int D{\bm \lambda} P[{\bm\lambda}] \sum_{i, j=1}^N {\rm e}^{-it |\lambda_i- \lambda_j|/2} \nonumber\\
& =\frac{1}{N^2} \sum_{i, j=1}^N\int d\lambda_{i}d\lambda_{j} R^{(2)}(\lambda_{i},\lambda_{j})  {\rm e}^{-it |\lambda_i- \lambda_j|/2}
\label{5}
\end{align}
where $R^{(2)}(\lambda_i,\lambda_j) = \int  P[\lambda] ...d\lambda_{i-1}d\lambda_{i+1}...d\lambda_{j-1}d\lambda_{j+1}...$ is the reduced JPDF.

Considering a GUE, represented by a set of ${N \times N}$ random matrix, the matrix elements are Gaussian random variables and follow a normal distribution with zero mean value and standard deviation of $1/\sqrt{N}$. Therefore, the energy levels of the system are confined to the interval $[-2, 2]$. Notably, prior researches have obtained the analytical form of reduced JPDF~\cite{mehta2004random,liu2018spectral},
\begin{equation}
\begin{aligned}
R^{(2)}(\lambda_i,\lambda_j) =\left\lbrace 
\begin{aligned}
  &\frac{\sqrt{4-\lambda_i^2}}{2 \pi}    &  i = j \\ 
  &\frac{N^2}{N(N-1)}\frac{\sqrt{\left( 4-\lambda_i^2\right) \left(4-\lambda_j^2 \right) }}{4\pi^2}\\
 &-\frac{N^2}{N(N-1)}\frac{\sin ^2 \left[N\left(\lambda_i-\lambda_j\right)\right]}{\left[N \pi\left(\lambda_i-\lambda_j\right)\right]^2}    & i \neq j
\end{aligned}
.\right.
\label{6}
\end{aligned}
\end{equation}

Upon the substitution of Eq. (\ref{6}) into Eq. (\ref{5}), we obtain the analytical expression of the GSFF for the case of two-point correlation. It is pointed out that the integral concerning $\lambda_{i}$ and $\lambda_{j}$ is independent on subindex $i,j$ in Eq.(\ref{5}). The summation is order of  ${\cal O}(N^{2})$ at early time. Altogether, the GSFF can be expressed as
\begin{equation}
\begin{aligned}
g^{(2)}\left( t\right) & = \mathcal{A}^{(2)}+\mathcal{B}^{(2)}\left( t\right) + \mathcal{C}^{(2)}\left( t\right)
\label{7} 
\end{aligned}
\end{equation}
where,
\begin{equation}
\begin{aligned}
\mathcal{A}^{(2)} = \frac{1}{N} \int \frac{\sqrt{4-\lambda_1^2}}{2 \pi} d \lambda_1 = \frac{1}{N};
\label{8} 
\end{aligned}
\end{equation}

\begin{equation}
\begin{aligned}
\mathcal{B}^{(2)}\left( t\right) = \int \frac{\sqrt{4-\lambda_1^2}}{2 \pi} \frac{\sqrt{4-\lambda_2^2}}{2 \pi} {\rm e}^{-i \frac{\left|\lambda_1-\lambda_2\right| t}{2}} d\lambda_1 d\lambda_2;
\label{9} 
\end{aligned}
\end{equation}

\begin{equation}
\begin{aligned}
\mathcal{C}^{(2)}\left( t\right) = -\int \frac{\sin ^2 \left[N\left(\lambda_1-\lambda_2\right)\right]}{\left[N \pi\left(\lambda_1-\lambda_2\right)\right]^2} {\rm e}^{-i\frac{\left|\lambda_1-\lambda_2\right| t}{2}} d\lambda_1 d\lambda_2.
\label{10} 
\end{aligned}
\end{equation}

Now we do the integral in Eq. (\ref{9}), the real part of which can be directly obtained,
\begin{align}
\text{Re}[\mathcal{B}^{(2)}\left( t\right)] & =\int \frac{\sqrt{4-\lambda_1^2}}{2 \pi} \frac{\sqrt{4-\lambda_2^2}}{2 \pi} \cos \left[ \frac{\left|\lambda_1-\lambda_2\right|t}{2}\right]  d\lambda_1 d\lambda_2\nonumber\\
& =\frac{4 J_1^2(t)}{t^2}.
\label{11}
\end{align}
where $J_1(t)$ is the Bessel function of the first kind. In the long time limit, the asymptotic behavior of $\text{Re}[\mathcal{B}^{(2)}\left( t\right)]$ reads ${8 \cos ^2\left(t+\frac{\pi}{4}\right)}/{\pi t^3}$, which is an oscillatory decay.

Nevertheless, a closed  analytical form of the  imaginary part: 
\begin{align}
\text{Im}[\mathcal{B}^{(2)}\left( t\right)]  = - \int &\frac{\sqrt{4-\lambda_1^2}}{2 \pi} \frac{\sqrt{4-\lambda_2^2}}{2 \pi}\nonumber\\
& \times\sin \left[ \frac{\left|\lambda_1-\lambda_2\right|t}{2}\right]  d\lambda_1 d\lambda_2
\label{12}
\end{align}
is hardly available. In the following discussion of $\text{Im}[\mathcal{B}^{(2)}\left( t\right)]$, we use numerics. It should be noticed that in the long-time limit, the integrand is strongly oscillating around zero. As such, $\text{Im}[\mathcal{B}^{(2)}\left( t\right)]$ decays to zero in the longtime limit, which takes the form of $-1/t$ in the long time limit.

The integral in  Eq. (\ref{10}) can be done by adopting the technique detailed in Ref.\cite{liu2018spectral}.  Specifically, we  introduce the variables $\mu=\lambda_1-\lambda_2 $ and $\nu = \lambda_2$, along with a cut-off $\nu\in \left[-\frac{\pi}{2},\frac{\pi}{2}\right]$ based on the Wigner semicircle law and distribution normalization, then we arrive at the following expressions for ${\cal C}^{(2)}(t)$
    \begin{equation}
    	\begin{aligned}
    		\text{Re}[\mathcal{C}^{(2)}\left( t\right)] & =\left\lbrace  
    		\begin{aligned}
    				&\frac{t-4N}{4N^2} & t<4N \\
    				& 0  & t \geq 4N\\		
    		\end{aligned}
    		;\right.
    	\end{aligned}
    \end{equation}
    and
    	\begin{align}
    		&\text{Im}[\mathcal{C}^{(2)}\left( t\right)]  =\nonumber\\
		&\left\lbrace  
    			\begin{aligned}
    				&\frac{t \log \left(\frac{16 N^2}{t^2}-1\right)+8N \coth ^{-1}\left(\frac{4 N}{t}\right)}{4 \pi  N^2} & t<4N \\
    				& 0 & t=4N\\
    				&\frac{t \log \left(1-\frac{16 N^2}{t^2}\right)+8 N \tanh ^{-1}\left(\frac{4 N}{t}\right)}{4 \pi  N^2} & t>4N \\
    			\end{aligned}
    		.\right.
    	\end{align}
    
Altogether, the sum of the real parts from the three terms leads to an analytical formulation of the real part of the two-level GSFF, expressed as:
    \begin{equation}
    	\begin{aligned}
    		\text{Re}[g^{(2)}\left( t\right)] & =\left\lbrace  
    		\begin{aligned}
    				&\frac{1}{N}+\frac{4 J_1^2(t)}{t^2}+\frac{t-4N}{4N^2} & t<4N \\
    				&\frac{1}{N}+\frac{4 J_1^2(t)}{t^2} & t \geq 4N\\		
    			\end{aligned}
    		.\right.
    	\end{aligned}
    \end{equation}
    
It can be checked that the analytical expression of  $\text{Re}[g^{(2)}\left( t\right)]$ in the above equation is consistent with the numerics of Eq.(\ref{2}). Specifically, the numerics can be done as follows:
\begin{align}
 \text{Re}[g^{(2)}\left( t\right)] 
 & = \text{Re} \left[ \int  \frac{d\epsilon}{N^2} \left\langle \sum_{i,j=1}^N \delta \left(\epsilon - \sigma \left(\lambda_i,\lambda_j\right)\right) \right\rangle_{\text{em}} {\rm e}^{-i \epsilon t} \right]\nonumber \\
 & =\text{Re} \left[ \frac{1}{N^2} \left\langle  \sum_{i,j=1}^N {\rm e}^{-i \frac{\left|\lambda_1-\lambda_2\right|}{2}t}  \right\rangle_{\text{em}}\right]\nonumber \\
 & = \frac{1}{N^2} \left\langle  \sum_{i,j=1}^N \cos\left[ \frac{\left|\lambda_1-\lambda_2\right|}{2}t\right]  \right\rangle_{\text{em}}.
\label{16}
\end{align}
where $\left\langle .\right\rangle_{\text{em}}$ denotes the ensemble average.
\begin{figure}[t]
    	 \centering 
    	 \includegraphics[width=\linewidth]{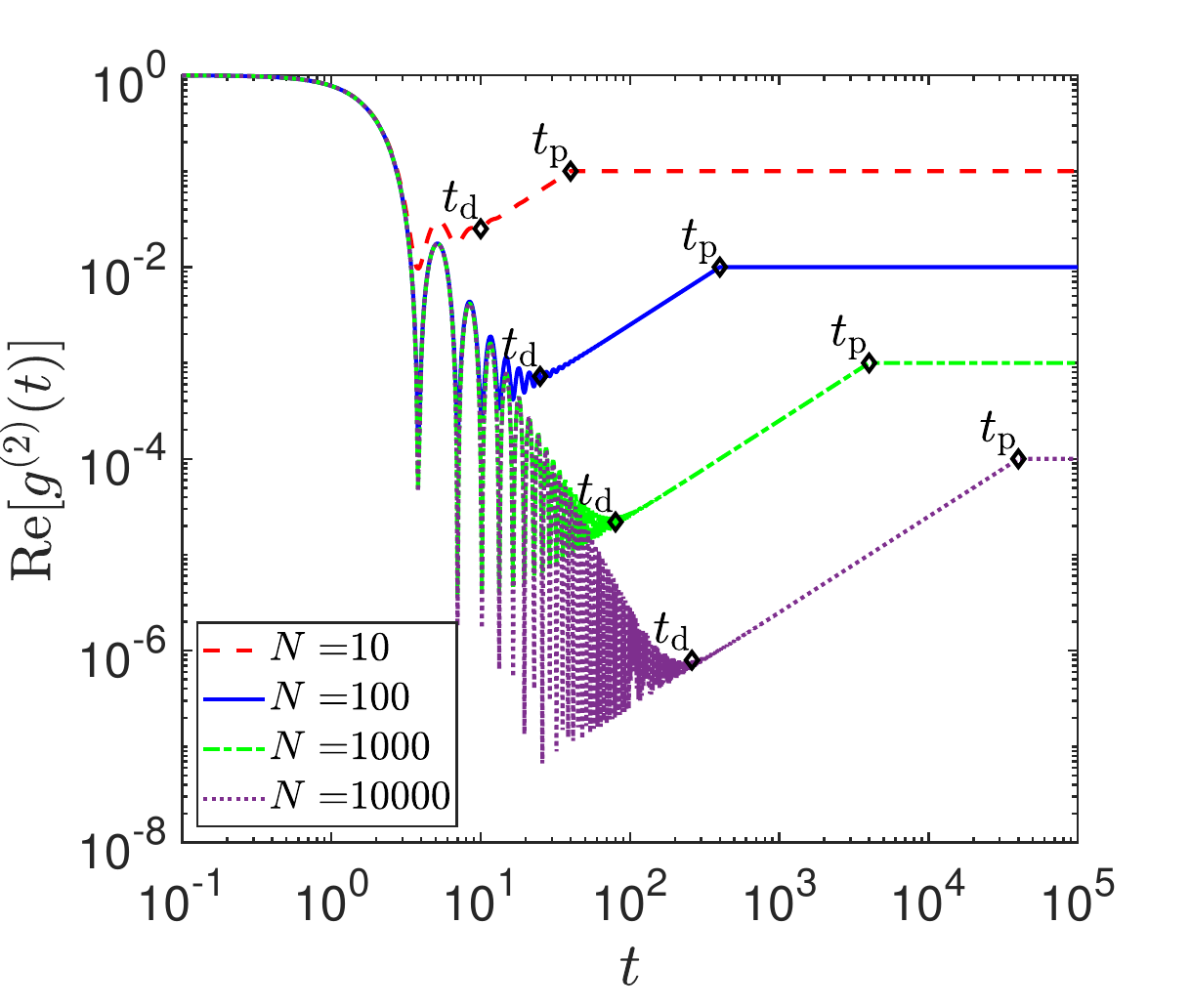}
	 \caption{Real part of the two-level generalized spectral form factor (GSFF) in Gaussian unitary ensembles (GUEs). Similar to SFF, the real part of the two-level GSFF shows a universal "dip-ramp-plateau" structure, as it is observed over a range of Hamiltonian dimensions $N=$10 (red dashed), 100 (blue solid),1000 (green dash-dotted), 10000 (purple dotted). $t_{\rm d}$ and $t_{\rm p}$ denote the time of the presence of dip and plateau, respectively.}
      \label{Fig1}
\end{figure}     

\subsection{Discussion of Real Part}
In Fig \ref{Fig1}, we present the real part of GSFF for the random matrix of varying dimension $N$. It is apparent that, across different $N$, the real part of the GSFF shows the same behaviors characterized by distinct features: a dip, an oscillation, a ramp, and a plateau. This universal pattern is referred to as the ``dip-ramp-plateau" structure, a hallmark of quantum chaotic systems~\cite{haake1991quantum, Saad2018ASR,ghar2018onset, Winer2020ER}. GSFF exhibits three distinct durations, $t<t_{\rm d}$, $t_{\rm d}<t<t_{\rm p}$ and $t>t_{\rm p}$, where $t_{\rm d}$ denotes the time of the presence of dip and $t_{\rm p}$ the time of the beginning of plateau. The time duration required for the real part of the GSFF to reach the plateau phase depends on the dimension of the system characterized by the time parameter $t_{\rm p}=4N$. These behaviors are consistent with the conventional SFF.  

\begin{figure}[t]
	 \centering 
	 \includegraphics[width=\linewidth]{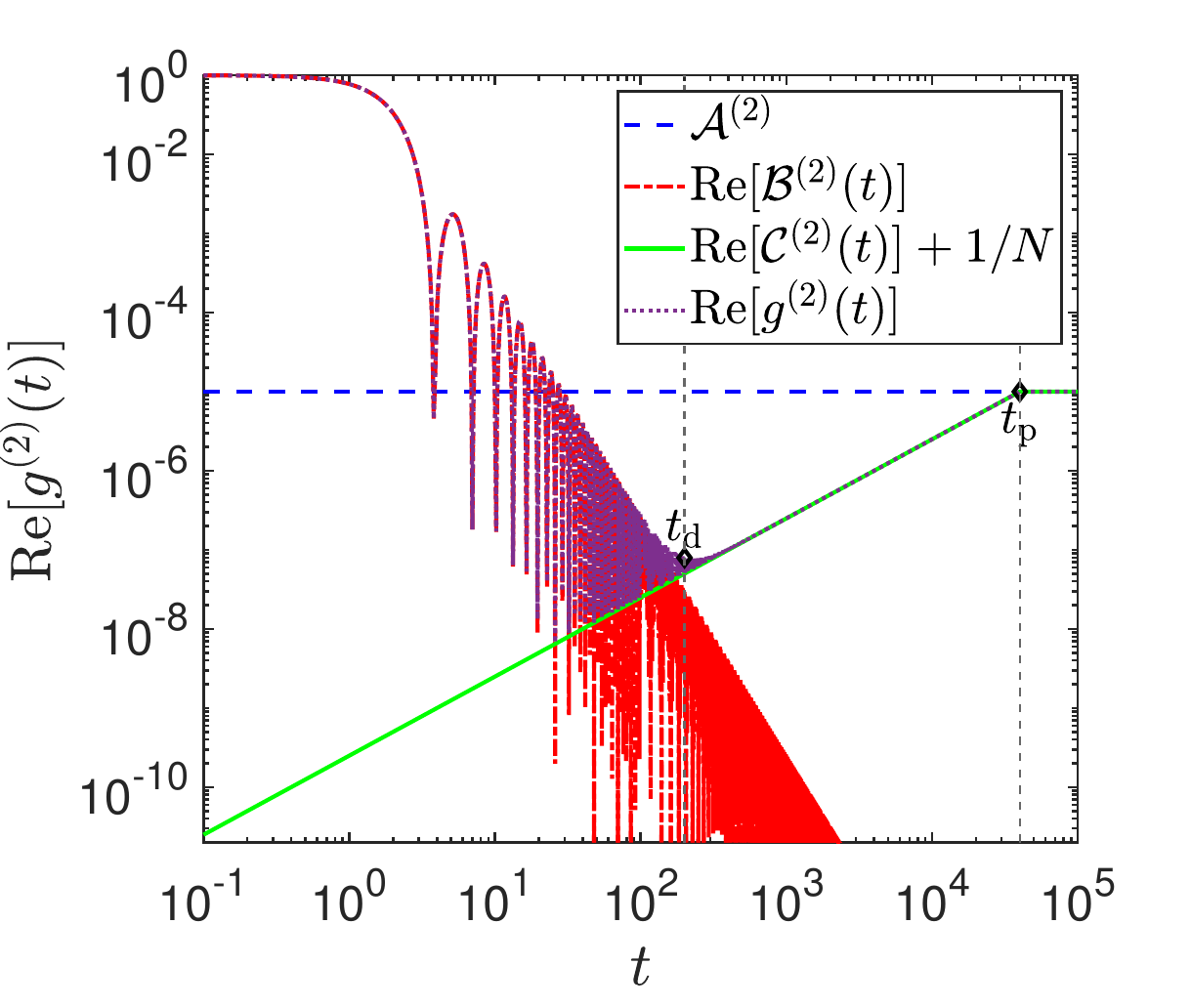}
      \caption{Comparative analysis of the real part of the two-level GSFF. We take the GUE with a dimension of $N=10000$ as an example. We show $\mathcal{A}^{(2)}$ (blue double dash), $\text{Re}[\mathcal{B}^{(2)}\left( t\right)]$ (red dash-dotted), $\text{Re}[\mathcal{C}^{(2)}\left( t\right)]+ 1/N $ (green solid), representing the real part of the results of the first, second, and the sum of the third and first terms of the integral defined in Eq. (\ref{7}). Additionally, we also show $\text{Re}[g^{(2)}\left( t\right)]$ (purple dotted) to encompass the entire real part, offering a comprehensive view of the contributions of individual integrals.}
	 \label{Fig2}
\end{figure}

To further distinguish the dominating contribution for each duration, we offer a comparative examination of the exact results and the contribution from an individual contribution for a specific case where $N=10000$, as shown in Fig \ref{Fig2}. The contribution of the first term denoted as $\mathcal{A}^{(2)}$ (blue dashed), the second term $\text{Re}[\mathcal{B}^{(2)}\left( t\right)]$ (red dash-dotted), and the cumulative impact of the first and third terms $\mathcal{A}^{(2)}+\text{Re}[\mathcal{C}^{(2)}\left( t\right)]$ (green solid) can be resolved from the curve of $\text{Re}[g^{(2)}\left( t\right)]$ as depicted in purple. In the first duration $t<t_{\rm d}$, GSFF shows oscillation, and our results reveal that the oscillation is mainly attributed to $\text{Re}[\mathcal{B}^{(2)}\left( t\right)]$. In the second duration $t_{\rm d}<t<t_{\rm p}$, GSFF ramps to a constant value. This behavior is attributed to $\text{Re}[\mathcal{C}^{(2)}\left( t\right)]$. In Fig.~\ref{Fig2}, we shift $\text{Re}[\mathcal{C}^{(2)}(t)]$ by amount of ${\cal A}^{(2)}$, i.e,
\begin{align}
\text{Re}[g^{(2)}\left( t\right)]\approx\frac{t}{4N^{2}},\quad t_{\rm d}<t<t_{\rm p}.
\label{17}
\end{align}
Notably, the slope of this ramp is $1/(4N^{2})$. As shown in Fig. \ref{Fig2}, the contribution from $\text{Re}[\mathcal{B}^{(2)}\left( t\right)]$ is much smaller than $\mathcal{A}^{(2)}+\text{Re}[\mathcal{C}^{(2)}\left( t\right)]$. So, in this duration, we can ignore the impact of $\text{Re}[\mathcal{B}^{(2)}\left( t\right)]$, as expressed in Eq. (\ref{17}). In the third duration, GSFF saturates to a constant value determined by ${\cal A}^{(2)}$. This term is exclusively associated with Wigner's semicircle law, and its plateau value reflects an inverse correlation with the dimension $N$. 

\begin{figure}[t]
   	   \centering
   	   \includegraphics[width=\linewidth]{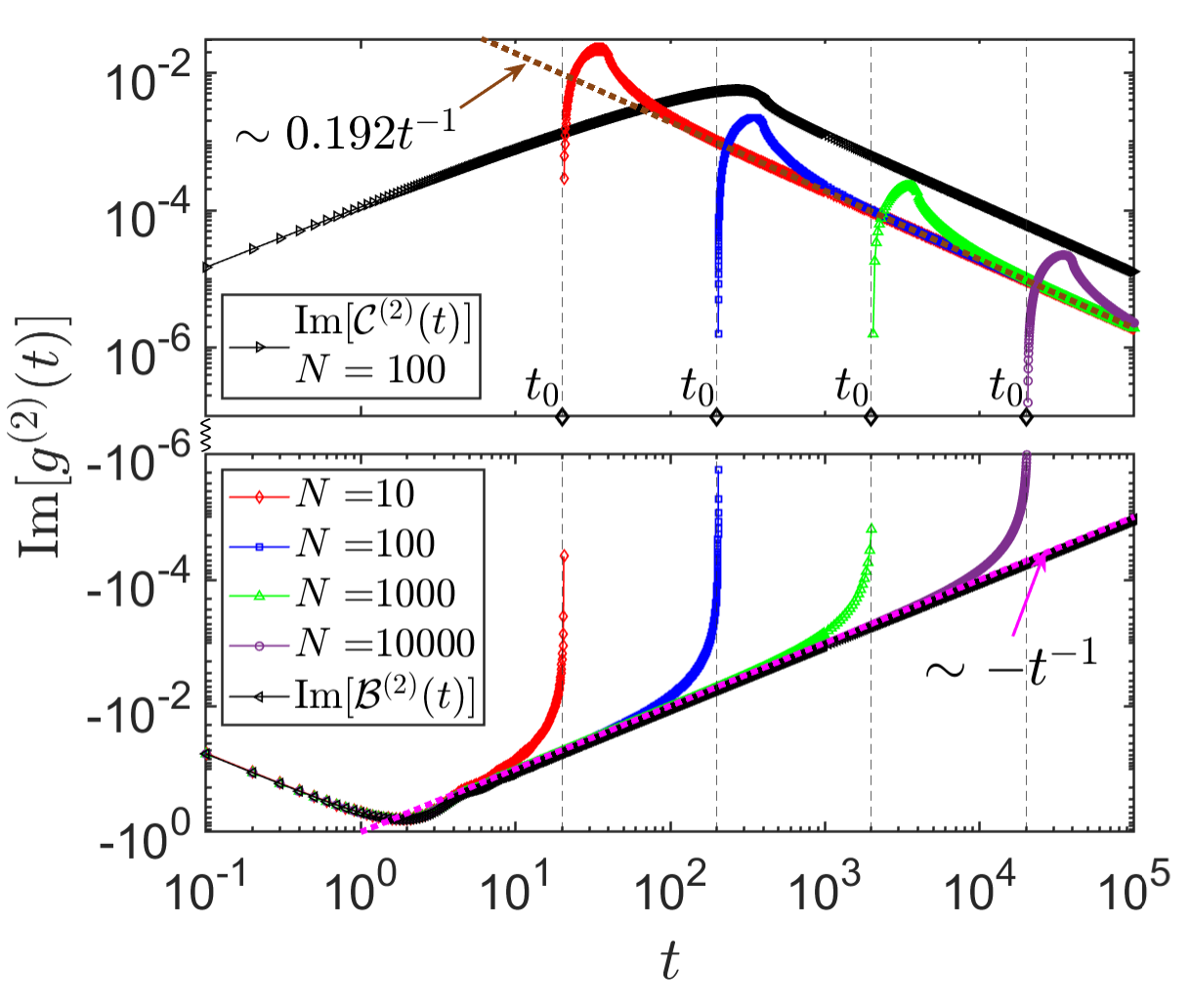}
        \caption{Imaginary part of two-level GSFF. The imaginary part presents a universal behavior at the early time and long time across various system dimensions $N=$10 (red diamond dots line), 100 (blue square dots), 1000 (green upper triangle dots), and 10000 (purple round dots). At the early time, the imaginary part $\text{Im}[g^{(2)}\left( t\right)]$ is primarily contributed by $\text{Im}[\mathcal{B}^{(2)}\left( t\right)]$(black triangle dots). At the long time, $\text{Im}[g^{(2)}\left( t\right)]$ is primarily contributed by $\text{Im}[\mathcal{C}^{(2)}\left( t\right)]$(black right triangle dots line). We take the system with $N=$100 as an example. The transition time $t_{0}$ (black diamond dots) for the imaginary part to change from negative to positive is determined by the system size, denoted as $t_{0} = 2N$ (black vertical dashed). In addition, at large $t$, $\text{Im}[g^{(2)}\left( t\right)]$ behaves as $t^{-1}$ (yellow dashed) and $\text{Im}[\mathcal{B}^{(2)}\left( t\right)]$ behaves as $-t^{-1}$ (magenta dashed).}
        \label{Fig3}
\end{figure}

\subsection{Discussion of Imaginary Part}
In contrast to the previous definition of SFF, a prominent new feature of GSFF is the presence of the imaginary part, which is shown in Fig. \ref{Fig3}. Remarkably, our results show the imaginary part presents a universal behavior at the early time and long time across various system dimensions ($N$). Specifically, the imaginary part initially exhibits an identical behavior, featuring the same dip in the region where the imaginary part is negative. Over time, each case manifests a peak in the region where the imaginary part is positive, then shares a power low decay of form $\sim0.192t^{-1}$. Another notable observation is the transition time $t_{0}$ for the imaginary part to change from negative to positive. Our results show that this time scale is determined by the system size, denoted as $t_{0} = 2N$.

The imaginary part of the GSFF is governed by different terms in the short time and long time limit. We take the case with $N=100$ as an example to illustrate which term dominates in Eq.(\ref{7}).  In the short time limit, $\text{Im}[\mathcal{B}^{(2)}\left( t\right)]$ is negative and $\text{Im}[\mathcal{C}^{(2)}\left( t\right)]$ is positive. But the latter is of two orders of magnitude smaller than the former. So it is evident that when $t < t_{0}$, the imaginary part $\text{Im}[g^{(2)}\left( t\right)]$ is primarily contributed by $\text{Im}[\mathcal{B}^{(2)}\left( t\right)]$, which is similar to the real part. As time progresses,  the magnitude of $\text{Im}[\mathcal{B}^{(2)}\left( t\right)]$ decrease, in a form of $\sim t^{-1}$ , and $\text{Im}[\mathcal{C}^{(2)}\left( t\right)]$ becomes the dominated one.
 Additionally, based on Eq.(\ref{8}) and Eq.(\ref{9}), when $t = 0$, $\text{Im}[g^{(2)}\left( t\right)] = 0$. 
\begin{figure}[t]
	   \centering
	   \includegraphics[width=\linewidth]{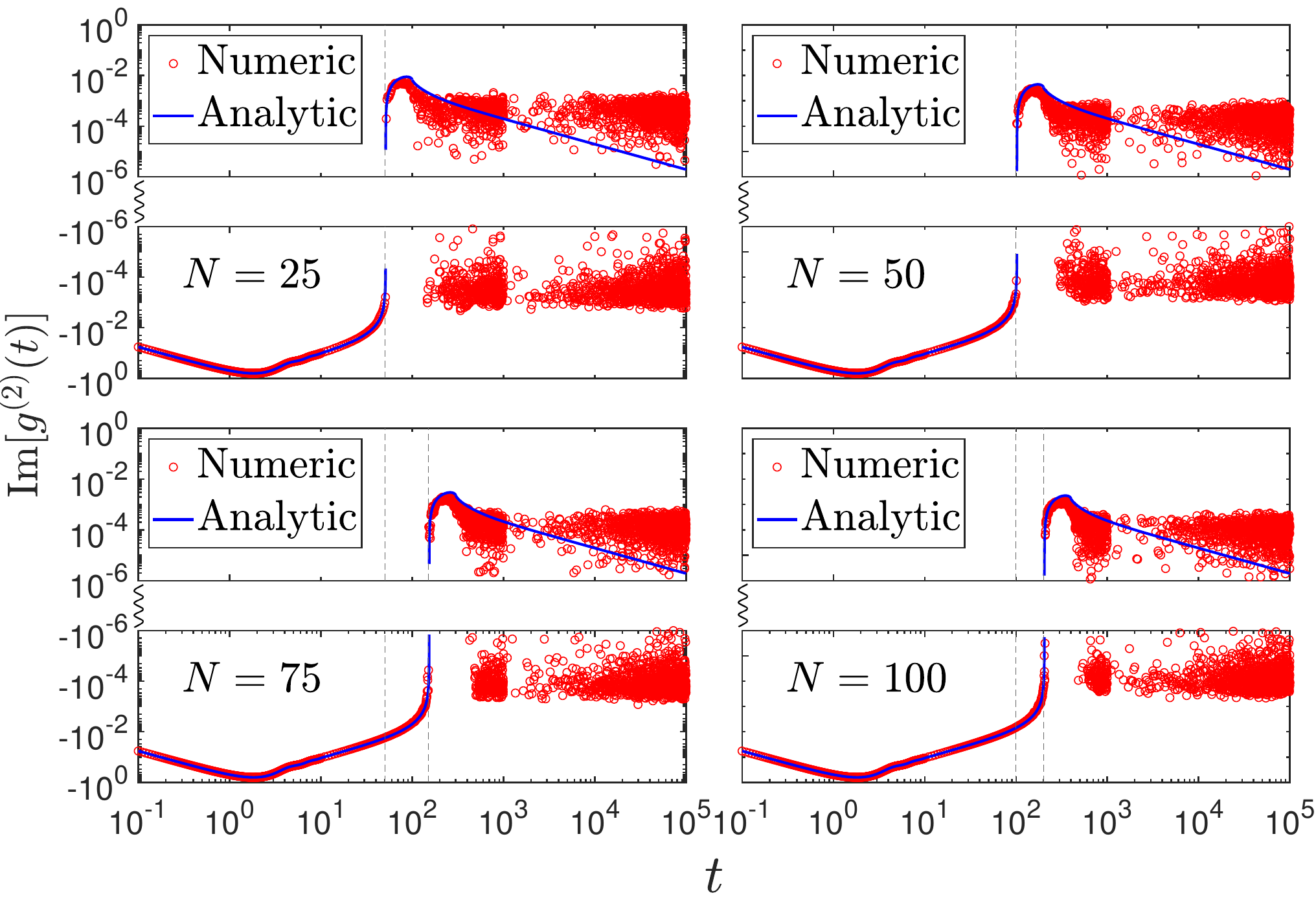}
        \caption{Comparative analysis of analytical and numerical results of the imaginary part for two-level GSFF. At the early time, the analytical results (blue solid) are consistent with the numerical results (red circle). In the later stage, the analytical results decay. The numerical results exhibit an oscillating behavior around 0 and a symmetric distribution. In numerical calculations, we used four ensembles with different system dimensions ($N$), in which $N=$ 25, 50, 75, and 100 respectively, and each ensemble comprises 5000 Hamiltonian matrices. The gray vertical dash lines represent the transition time $t_{0}$ of the imaginary part in different systems.}
         \label{Fig4}
\end{figure}

To verify our previous analysis, a numerical statistics approach is utilized to derive the imaginary component of GSFF. The procedure is formulated as follows:
\begin{align}
\text{Im} [g^{(2)}\left( t\right)]& = \text{Im} \left[ \int  \frac{d\epsilon}{N^2} \left\langle \sum_{i,j=1}^N \delta \left(\epsilon - \sigma \left(\lambda_i,\lambda_j\right)\right) \right\rangle_{\text{em}} {\rm e}^{-i \epsilon t} \right]\nonumber \\
 & =\text{Im} \left[ \frac{1}{N^2} \left\langle  \sum_{i,j=1}^N {\rm e}^{- \frac{i\left|\lambda_1-\lambda_2\right|}{2}t}  \right\rangle_{\text{em}}\right] \nonumber\\
 & = \frac{1}{N^2} \left\langle  -\sum_{i,j=1}^N \sin\left[ \frac{\left|\lambda_1-\lambda_2\right|}{2}t\right]  \right\rangle_{\text{em}}.
\label{18}
\end{align}
where $\langle.\rangle_{\rm em}$ denote ensemble average.  Four ensembles with different dimensions are generated, each consisting of 5000 random matrices. The dimensions of the random matrices for each system are $N=$25, 50, 75, and 100, respectively. The results for the imaginary part of GSFF, obtained through numerical statistics, are presented as the red round dots in Fig.~\ref{Fig4}, while the analytical result is shown as the blue solid line. When $t$ is small, $\text{Im}[g^{(2)}\left( t\right)]$ is negative and forms a dip, consistent with the analytical result. When $t$ is large, $\text{Im}[g^{(2)}\left( t\right)]$ exhibits violent oscillation near $0$.  This means that even small changes in energy level differences can lead to drastic changes in function value. Besides, Fig.~\ref{Fig4} reveals a symmetrical distribution of the oscillation around 0, indicating that the imaginary part average is 0. This is in line with the power law decay exhibited by the analytical result over an extended duration. Additionally, comparing systems with varying dimensions indicates that as $N$ increases, the region for the numerical results to be consistent with the analytical results extends, as indicated by the gray lines. Therefore, it can be inferred that the oscillation behavior of numerical results over an extended duration is a finite-size effect. The imaginary part of GSFF demonstrates a decay towards zero over a long time, which is different from the plateau of the real part in a long time.

Overall, the novel insights into the properties of these behaviors, particularly the behavior of the imaginary part, illuminate the complex and fascinating characteristics of GSFF. These findings, exemplified by their intriguing time dependencies and statistical correlations, enrich our understanding of energy level statistics and the real and imaginary parts of GSFF. This analysis can be extended to higher-level correlation.

\section{Numerical Statistics of Three-level Generalized Spectral Form Factor}
\label{sec:4GSFF}
\begin{figure}[t]
   	\centering
   	\includegraphics[width=\linewidth]{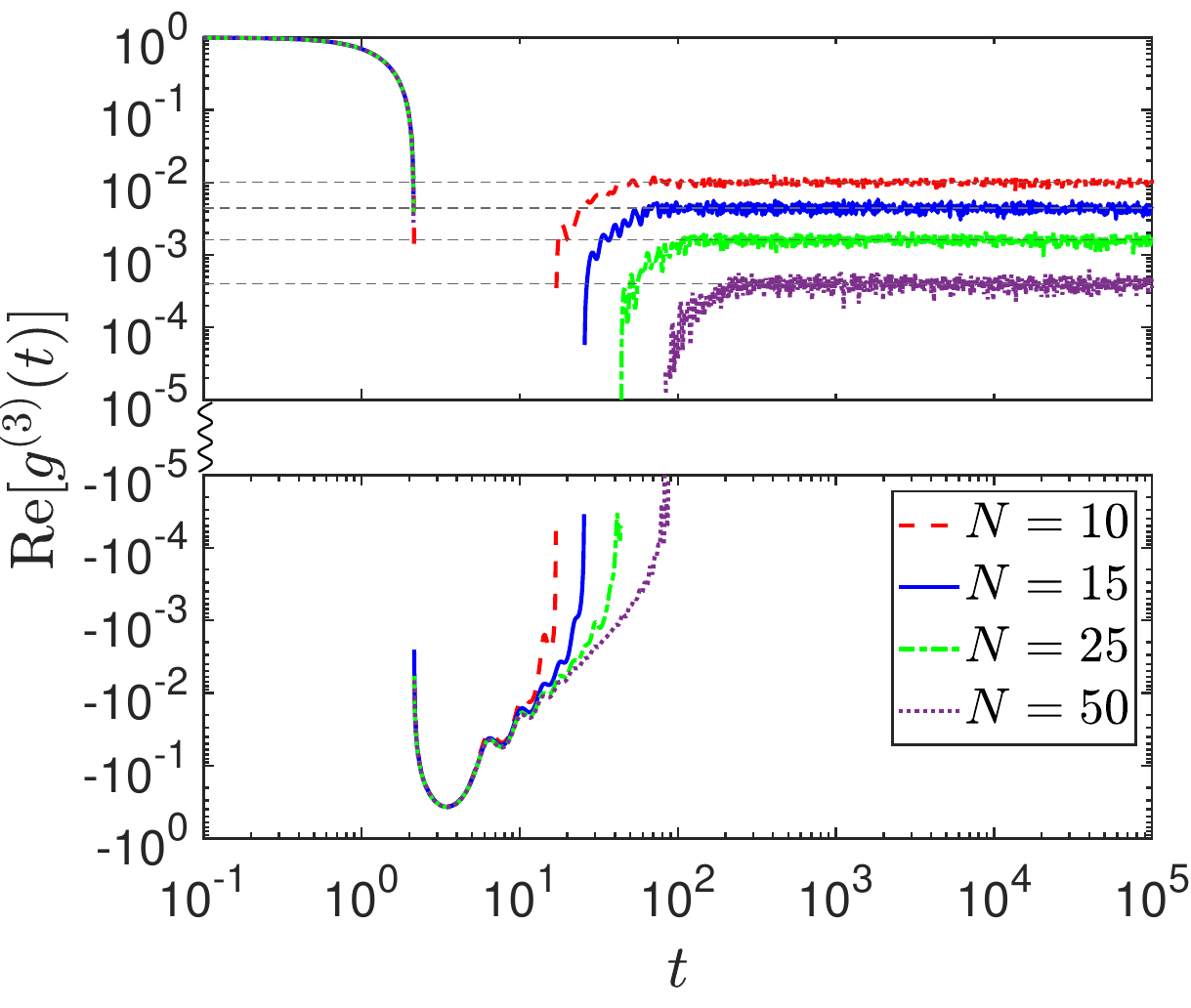}
     \caption{Numerical results of the real part of three-level GSFF. In numerical calculations, we used four ensembles with different system dimensions ($N$), in which $N=$ 10 (red double dash), 15 (blue solid), 25 (green dash-dotted), and 50 (purple dotted) respectively, and each ensemble comprises 5000 Hamiltonian matrices. The real part of three-level GSFF $\text{Re}[g^{(3)}\left( t\right)]$ of systems of different dimensions show similar behavior. In the long time limit, the real part shows a plateau. The value of the plateau is related to the dimension $N$, expressed as $\text{ Re}[g^{(3)}\left( t_{\rm large}\right)] = 1/N^2$ (gray horizontal dashed lines).}
   	\label{Fig5}
\end{figure}

Let us direct our focus towards the three-level GSFF, also specifically in the context of GUE governed by a set of random matrix $\left\lbrace M_{N \times N}\right\rbrace$. In this case, the analytical expression is hardly available. In our following discussion, we use numerical results. Similar to the previous choice, there are four ensembles, and each ensemble contains 5000 random matrices. The dimensions of the matrices corresponding to each ensemble are $N=$10,15,25,50 respectively.  The representation of the three-level GSFF by numerical method can be expressed as follows:
\begin{align}
g^{(3)}\left( t\right) & =\int c^{(3)}\left( \epsilon\right) {\rm e}^{-i \epsilon t} d \epsilon\nonumber \\
& = \int \frac{d \epsilon}{N^3}\left\langle \sum_{i,j,k=1}^N \delta\left[\epsilon- \sigma \left(\lambda_i,\lambda_j,\lambda_k\right)\right] \right\rangle_{\text{em}} {\rm e}^{-i \epsilon t} \nonumber \\
& = \left\langle \int \frac{d \epsilon}{N^3} \sum_{i,j,k=1}^N \delta\left[\epsilon- \sigma \left(\lambda_i,\lambda_j,\lambda_k\right)\right] {\rm e}^{-i \epsilon t} \right\rangle_{\text{em}}\nonumber\\
& =\frac{1}{N^3} \left\langle  \sum_{i,j,k=1}^N {\rm e}^{-i \sigma \left(\lambda_i,\lambda_j,\lambda_k\right) t} \right\rangle_{\text{em}},
\label{19}
\end{align}
where $\sigma \left(\lambda_i,\lambda_j,\lambda_k\right)$ represents the standard deviation of three energy levels. Upon numerical calculations, the real parts $\text{Re}[g^{(3)}\left( t\right)]$ and imaginary parts $\text{Im}[g^{(3)}\left( t\right)]$ of the GSFF can be extracted. The results of the real part of GSFF, $\text{Re}[g^{(3)}\left( t\right)]$, are shown in Fig.~\ref{Fig5}, which can be divided into three time durations. When $t$ is small, $\text{Re}[g^{(3)}\left( t\right)]$ is positive, decreasing from unity, and the behavior of the real part is completely consistent across systems of varying sizes($N$). In the second region, a negative dip appears in $\text{Re}[g^{(3)}\left( t\right)]$. In this region, as time increases, for different $N$, the real parts show differences after experiencing a small oscillation, and the behavior is no longer completely consistent. Moreover, it can be found that the larger $N$ is, the later the real part goes from the negative to the positive.  As $t$ further increases, the real parts enter the third region and become positive again. The behavior of the real part shows a plateau, and the value of the plateau is related to the size of the system. This is qualitatively similar to the two-level GSFF. Specifically, $\text{Re}[g^{(3)}\left( t_{\rm large}\right)] = 1/N^2$.
\begin{figure}[t]
	   \centering
	   \includegraphics[width=\linewidth]{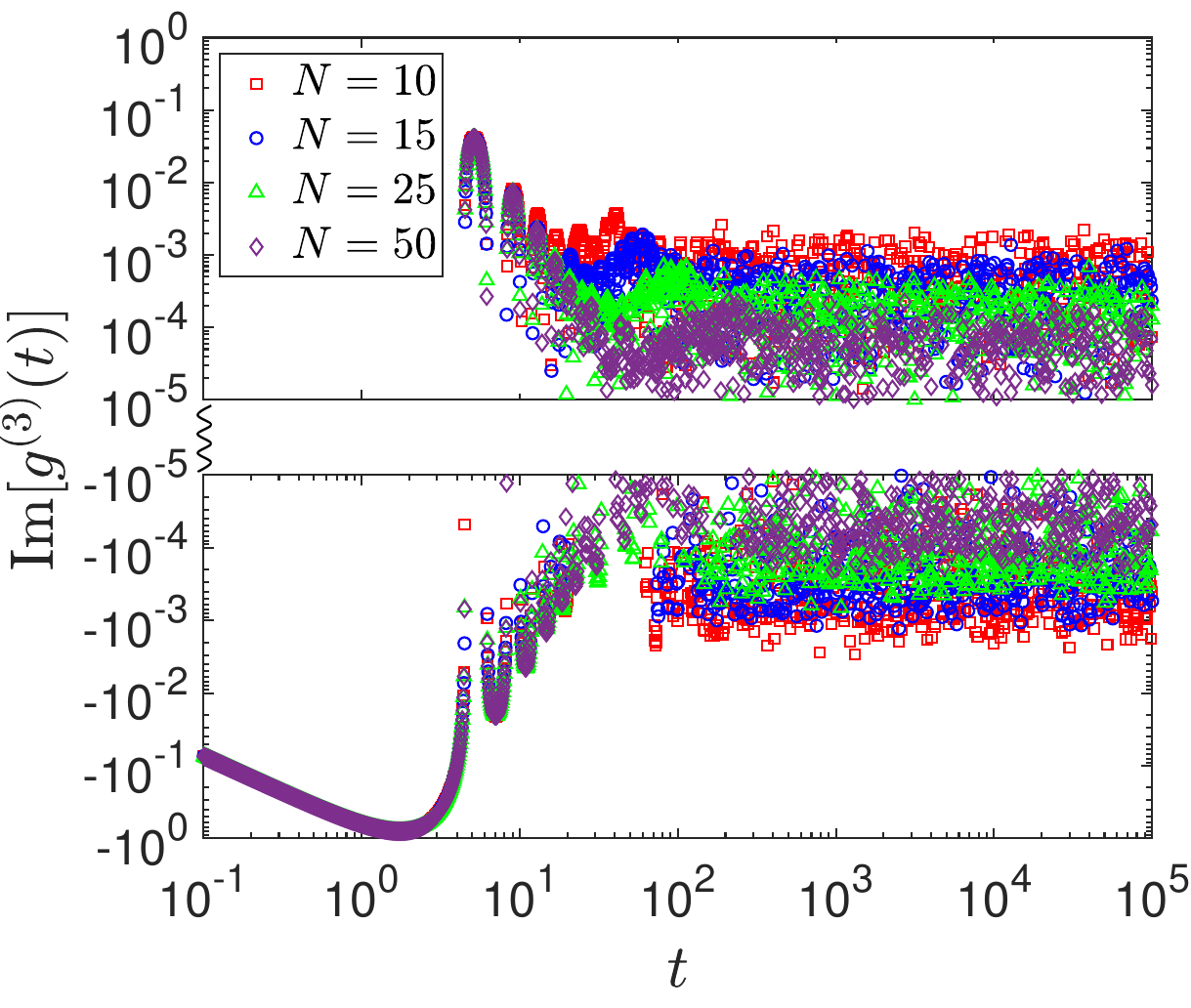}
	   \caption{Numerical results of the imaginary part of three-level GSFF. In numerical calculations, we used four ensembles with different system dimensions ($N$), in which $N=$ 10 (red square dots), 15 (blue round dots), 25 (green upper triangle dots), and 50 (purple diamond dots) respectively, and each ensemble comprises 5000 Hamiltonian matrices. The imaginary part of three-level GSFF $\text{Im}[g^{(3)}\left( t\right)]$  of systems of different dimensions show similar behavior. The behavior of $\text{Im}[g^{(3)}\left( t\right)]$ has a common dip and an oscillatory decay then oscillates around 0, which can be considered a finite-size effect.}
         \label{Fig6}
\end{figure}

In Fig.~\ref{Fig6}, we present the results of the imaginary part $\text{Im}[g^{(3)}\left( t\right)]$, which behaves similarly to the counterpart of the two-level GSFF. It is notable that, based on Eq.(\ref{19}), $\text{Im}[g^{(3)}\left( 0\right)] = 0$. Therefore, when $t$ is very small, $\text{Im}[g^{(3)}\left( t\right)]$ starts to decrease from 0, and for systems of different sizes, the behaviors of the imaginary part are the same.  A dip appears in the negative region. As $t$ increases, the behavior of $\text{Im}[g^{(3)}\left( t\right)]$ indicates a oscillatory  decay then oscillates around 0.
In addition, we observed that the oscillation amplitude of the long-time behavior of $\text{Im}[g^{(3)}\left( t\right)]$ attenuates as $N$ increases. This implies that the imaginary part tends to 0 in the thermal dynamical limit. Therefore, the oscillatory behavior can also be considered a finite-size effect.

\section{Summary}
\label{summary}
In summary, we extend the concept of SFF to incorporate the high-order level correlation in the spectrum. 
The GSFF is obtained by Fourier transforming the energy level correlation function that is defined by the standard deviation of energy levels. In contrast to conventional approaches that only examine the differences between two energy levels, our method enables a comprehensive exploration of correlations across arbitrary energy levels. 

In contrast to SFF, GSFF is complex. In the context of GUE, when considering the correlation of two energy levels, the real part of the GSFF aligns with the result of SFF.  It shows the dip-ramp-plateau structure, which is consistent with the SFF. Moreover, we extend our analysis to the imaginary part of the GSFF, revealing structural features characterized by dip in the negative region and universal decay in the positive region.  After comparative analysis with the results of numerical calculations, we find that the imaginary part of GSFF decays to zero in a long time limit.  Additionally, for the GSFF in scenarios involving three energy level correlations by numerics, the real part has a dip and a plateau structure, and the imaginary part has a dip, an oscillatory decay, and a long-time oscillation feature. 
It is worth noting that the characteristic times associated with spectral structures of two-level and three-level correlations are intimately linked to the system size ($N$). In short, the GSFF offers a robust computational approach for assessing energy level correlations at higher orders, enabling a deeper understanding of the energy level distribution of the system.

\textit{Acknowledgement}. We want to thank Tian-Gang Zhou for the helpful discussion. This work is supported by NSFC 12174300, Innovation Program for Quantum Science and Technology (Grants No.2021ZD0302001), the Fundamental Research Funds for the Central Universities (Grant No. 71211819000001) and Tang Scholar.


%

\end{document}